\documentclass[prl,aps,epsfig,superscriptaddress]{revtex4}

\usepackage{amssymb}
\usepackage{mathtools}
\usepackage[linkcolor=blue,anchorcolor=black,citecolor=blue]{hyperref}
\usepackage{graphicx}

\begin{document}
	
\title{Experimental Hamiltonian Learning of An 11-qubit Solid-State Quantum Spin Register}

\author{P.-Y. Hou\textsuperscript{*}, L. He\textsuperscript{*}, F. Wang\textsuperscript{*}, X.-Z. Huang, W.-G. Zhang, X.-L. Ouyang, X. Wang, W.-Q. Lian, X.-Y. Chang, L.-M. Duan\textsuperscript{$\dagger$}}

\affiliation{Center for Quantum Information, Institute for Interdisciplinary Information Sciences, Tsinghua University, Beijing 100084, China}

\begin{abstract}
	Learning Hamiltonian of a quantum system is indispensable for prediction of the system dynamics and realization of high fidelity quantum gates. However, it is a significant challenge to efficiently characterize the Hamiltonian when its Hilbert space dimension grows exponentially with the system size. Here, we experimentally demonstrate an adaptive method to learn the effective Hamiltonian of an 11-qubit quantum system consisting of one electron spin and ten nuclear spins associated with a single Nitrogen-Vacancy center in a diamond. We validate the estimated Hamiltonian by designing universal quantum gates based on the learnt Hamiltonian parameters and demonstrate high-fidelity gates in experiment.
Our experimental demonstration shows a well-characterized 11-qubit quantum spin register with the ability to test quantum algorithms and to act as a multi-qubit single node in a quantum network.
\end{abstract}

\maketitle

\section*{Introduction}

Experimental realization of quantum registers is an essential task for quantum information processing.
Lots of experiments towards this goal have been implemented on different physical systems such as trapped ions\cite{Monz11, Friis18, Zhang17}, solid-state spins\cite{Andreas16,Humphreys18,Childress13,Dutt07,Pfaff,Van17,Abobeih18,Taminiau12,Taminiau14,Goldman15,Casanova17}, neutral atoms\cite{Bernien17} and superconducting qubits\cite{Kelly15,IBM16}.
A common problem among several physical systems is the crosstalk between multiple qubits during individual addressing which induces state error to other qubits and reduces the gate fidelity.
To overcome this problem, a way is to learn and fully characterize the many-body coupling Hamiltonian\cite{Burgarth01, Burgarth02, Burgarth03, Franco09, Zhang14, Silva11, Wiebe14, wang2015hamiltonian, senko2014coherent} which determines the system dynamics and crosstalk interaction.
With the knowledge of the whole system Hamiltonian, one can predict the evolution of any initial states and design optimized gate operations to reduce crosstalk errors and achieve high-fidelity gates in a multi-qubit register.
Hamiltonian can often be fully described by some essential parameters, which characterize the coupling between the qubits. The number of these parameters typically only scales polynomially with the system size.

We focus on a solid-state spin system associated with a single nitrogen vacancy (NV) center in a bulk diamond, which consists of one electron spin and multiple surrounding nuclear spins. Such a system has been demonstrated to be a promising platform for quantum network \cite{Andreas16,Humphreys18,Childress13,Dutt07,Pfaff,Van17}, quantum information processing\cite{Taminiau14,Cramer16}, and quantum sensing\cite{Sense1,Sense2,Sense3,Sense4,Sense5,Sense6}.
Recently, dynamical decoupling technique has been well developed on this system to significantly extend the electron spin coherence time and to implement universal control of surrounding nuclear spins\cite{Taminiau14,Andreas16,Abobeih18,Taminiau12}.
Moreover, most surrounding nuclear spins can be further individually polarized and read out by such nuclear spin gates implemented by manipulating the NV electron spin via an optimized dynamical decoupling sequence.

For this solid-state spin register, the interactions among all the spin qubits are constantly on, which makes it inevitable to have crosstalk error from the other nuclear spins while controlling the target nuclear spin.
To mitigate the gate error due to such crosstalk, it is worthwhile to precisely characterize the system Hamiltonian. Based on the learnt Hamiltonian parameters, one can then optimize the dynamical decoupling sequence to minimize the effect from the unwanted crosstalk couplings for implementation of a universal set of quantum gates.
Moreover, NV center systems can be remotely entangled through photonic links \cite{bernien13,Pfaff,kalb17}, which provides a possibility to realize a scalable quantum network. A well-characterized multi-qubit quantum spin register can be used as an effective quantum node for such a quantum network.

In this paper, we experimentally characterize the effective many-body coupling Hamiltonian of an NV center system containing one electron spin and ten weakly coupled $^{13}C$ nuclear spins.
We first probe the surrounding spin environment by applying dynamical decoupling sequence on the electron spin to obtain a spectroscopy. 
In the dynamical decoupling spectroscopy, we identify $10$ dominant nuclear spins which give strong signals to the electron spin coherence. The essential Hamiltonian parameters for the target system are the hyperfine interactions between the NV electron spin and the $10$ nuclear spins. We first roughly extract these hyperfine interaction parameters by fitting the simulated data to the experimental spectroscopy.
Then, we precisely characterize the hyperfine parameters of each nuclear spin by measuring the  nuclear Larmor frequencieswith different electron spin states. We apply an adaptive method\cite{Bonato15,Cappellaro12} based on the quantum phase estimation algorithm in a sequence of Ramsey-interferometry experiments to improve the efficiency of the frequency measurement.
To validate the estimated Hamiltonian parameters, we numerically optimize quantum gates based on the learnt Hamiltonian parameters and experimentally demonstrate a universal set of quantum gates for this $11$-qubit system of one electron spin and ten nuclear spins with high gate fidelities.

\section*{Results}

\subsection{System Description}
We perform the experiment on a type-IIa CVD synthetic diamond sample with the natural abundance of $^{13}C$ ($\sim 1.1\%$) that has a nuclear spin $I=1/2$ ($|m_{I}=\frac{1}{2}\rangle \equiv |\uparrow\rangle, |m_{I}=-\frac{1}{2}\rangle \equiv |\downarrow\rangle$). The negative charge state of an NV center has an electron spin $S=1$ ($|m_{s}=0\rangle \equiv |0\rangle, |m_{s}=\pm1\rangle \equiv |\pm1\rangle$) which can be optically initialized with a fidelity over $99\%$ through the intersystem crossing\cite{Goldman15} and read out with an average fidelity of $90\%$ ($F_{b}=81\%$ for the bright state and $F_{d}=99\%$ for the dark state )) in a single shot at cryogenic temperature.\ With a magnetic field $B_{z}$ along the NV symmetry axis, the system Hamiltonian of the NV center is described by
\begin{equation}\label{key}
\hat{\mathbf{H}}= \Delta\hat{S_{z}}^{2}+ \gamma_{e}B_{z}\hat{S}_{z}+ \displaystyle{\sum_{i}^{}(\hat{\mathbf{S}}\cdot\hat{\mathbf{A}}_{i}\cdot\hat{\mathbf{I}}_{i}+\gamma_{n}B_{z} \hat{I}_{i,z})}
\end{equation}
where the NV symmetry axis is defined as the $z$ axis. The electron (nuclear) spin operator $\hat{\mathbf{S}}$ ($\hat{\mathbf{I}}$) contains the Pauli matrixes $\hat{S}_{x}$, $\hat{S}_{y}$, $\hat{S}_{z}$ ($\hat{I}_{x}$, $\hat{I}_{y}$, $\hat{I}_{z}$);
$\Delta$ is the zero-field splitting of 2.8776 $GHz$; $\gamma_{e}$ ($\gamma_{n}$) is the gyromagnetic ratio of the electron spin ($^{13}C$ nuclear spin); $\hat{\mathbf{A}}_{i}$ is the tensor of hyperfine interaction between the electron spin and the nuclear spin $i$.
Dipole-dipole interactions between nuclear spins are typically negligible.

In the rotating frame defined by the Hamiltonian $H_0=\Delta\hat{\mathbf{S}}^{2}+\gamma_{e}B \hat{S}_{z}$, by neglecting the fast oscillation terms, we get the effective Hamiltonian, which is described by
\begin{equation}\label{key}
\hat{\mathbf{H}}_{eff}= \displaystyle{\sum_{i}^{}(A_{i,zx}\hat{S}_{z}\hat{I}_{i,x}+A_{i,zy}\hat{S}_{z}\hat{I}_{i,y}+A_{i,zz}\hat{S}_{z}\hat{I}_{i,z}+\gamma_{n}B_{z}\hat{I}_{i,z})}=\displaystyle{\sum_{i}^{}\hat{H}_{i}}
\end{equation}
In Equation (2), the effective Hamiltonian $\hat{\mathbf{H}}_{eff}$ equals the sum of the subsystem Hamiltonians $\hat{H}_{i}$ which describes the hyperfine coupling between the electron spin and each nuclear spin.
In particular, $\hat{H}_{i}$ can be simplified as $\hat{H}^{'}_{i}=A^{'}_{i,zx}\hat{S}_{z}\hat{I}_{i,x}+A^{'}_{i,zz}\hat{S}_{z}\hat{I}_{i,z}+\gamma_{n}B_{z}\hat{I}_{i,z}$ by redefining the $x$ axis for each nuclear spin so that $A^{'}_{zy}=0$ (the $x,y$ axes for different nuclear spins can be defined independently as we have ignored the direct coupling terms between the nuclear spins). In the following, to simplify notation, we denote $A^{'}_{zz}$ and $A^{'}_{zx}$ in the rotated frame still as $A_{zz}$ and $A_{zx}$ by setting $A^{'}_{zy}=0$.
Therefore, the main task of learning the whole system Hamiltonian is simplified to characterization of the hyperfine parameters $\{A_{zz},A_{zx}\}$ for all the nuclear spins.

\begin{figure*}[t]
	\includegraphics[width=150mm]{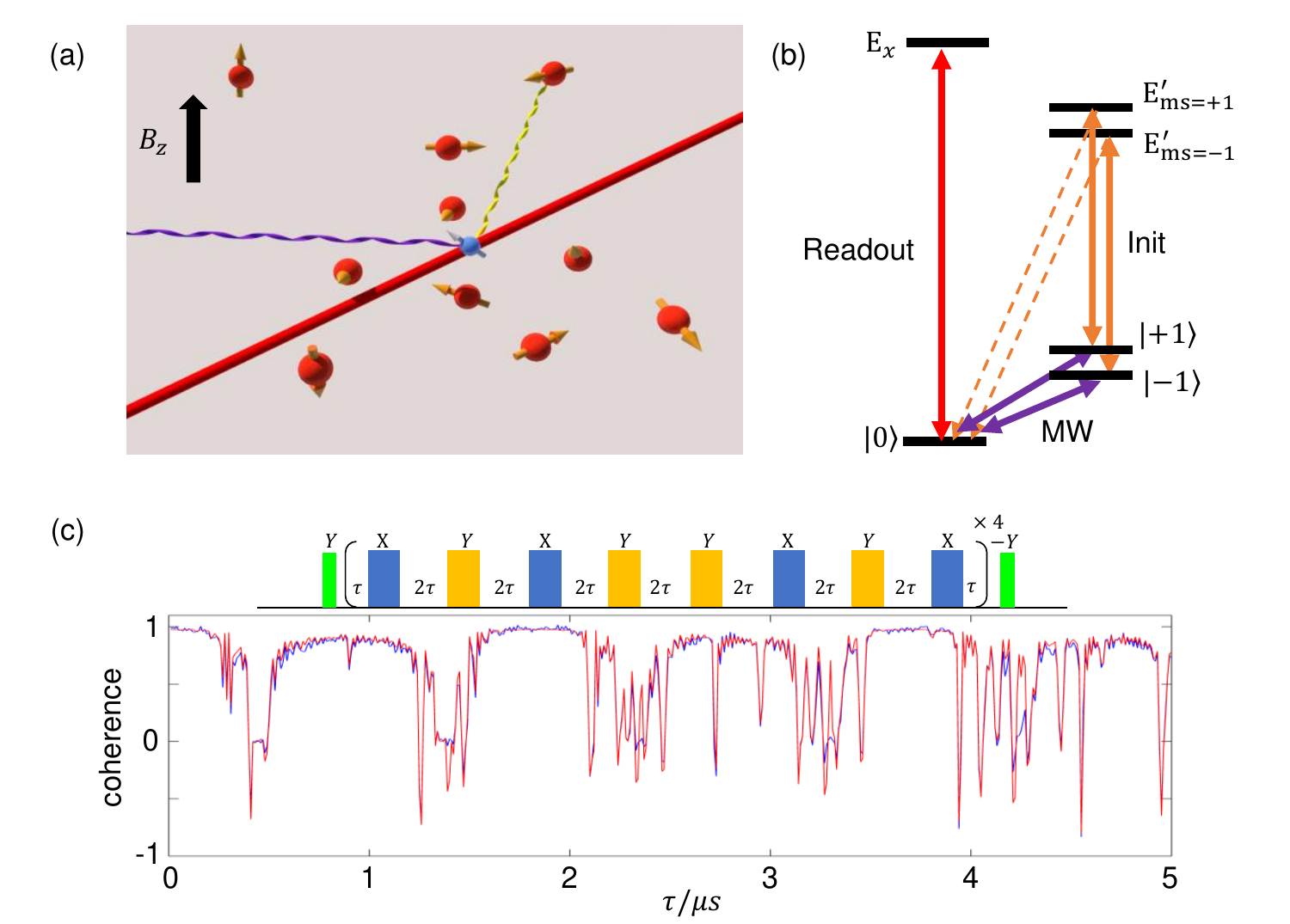}
	\caption{(a) Illustration of an NV center system consisting of one electron spin (blue ball) and multiple nuclear spins (red balls). The electron spin is optically initialized and read out by the resonant laser (red line), manipulated by microwave fields (purple wave). (b) Diagram of relevant energy levels of the NV center, where $|0\rangle$ ($|\pm1\rangle$) denotes respectively the bright (dark) state under a readout laser, and they are coherently manipulated by microwave signals. The state $|\pm1\rangle$ can be optically pumped to $|0\rangle$ by an initialization laser. (c) Dynamical decoupling spectroscopy probed by the electron spin with a CPMG-32 pulse sequence. Blue and red lines denote the experimental data and the simulation results for the $10$ resolved nuclear spins, respectively.\label{fig:fig1}}
\end{figure*}

\subsection{Identify nuclear spins from dynamical decoupling spectroscopy}
To explore the spin environment of an NV center, we prepare the electron spin in a superposition $(|0\rangle+|-1\rangle)/\sqrt{2}$ with a $\pi/2$ pulse and apply a Carr-Purcell-Meiboom-Gill (CPMG) type of dynamical decoupling sequence with $32$ $\pi$-pulses.
The CPMG sequence is formed by N $\pi$-pulses with the configuration of $(\tau-\pi-\tau)^{\times N}$, where $2\tau$ is the interval between two neighboring $\pi$-pulses. The phase of $\pi$-pulses in the CPMG sequence follows the XY-8 scheme shown in Fig.\ref{fig:fig1}(c).
The final state is projected on the $z$ basis by a second $\pi/2$ pulse and is read out with optical detection. By scanning the interval time in the range of $0<\tau<50\mu s$, we obtain the dynamical decoupling spectroscopy which shows the electron spin coherence as a function of the time $\tau$.
Ten nuclear spins that give strong dip signals to the electron spin coherence are resolved from the dynamical decoupling spectroscopy using the method in Ref. \cite{Taminiau12}.
By fitting the simulation results to the experimental spectroscopy, the hyperfine parameters $\{A_{zz},A_{zx}\}$ of $10$ resolved nuclear spins are extracted with limited precision. This precision is limited because all the nuclear spins, including the unresolved ones, contribute collectively to the dynamical decoupling spectroscopy, and it is hard to estimate the hyperfine parameters of each nuclear spin individually.
These extracted hyperfine parameters, although with limited precision, still allow us to perform quantum gates of limited fidelities on the electron and the nuclear spins. Through these gates, we are able to roughly polarize, control and read out these resolved nuclear spins.
Initially, these $10$ nuclear spins can be polarized with the fidelities ranging from $55\%$ to $85\%$.

\subsection{Precise characterization of the hyperfine parameters by adaptive quantum phase estimation}
Nuclear spin Larmor precession frequency is affected by the hyperfine interaction with the electron spin.
For a weakly coupled $^{13}C$ nuclear spin,  $f_{\pm}$, defined as the precession frequency of the nuclear spin when the electron spin is in the $|\pm1\rangle$ state, is given by
\begin{equation}
f_{\pm} = \frac{1}{2\pi}\sqrt{A_{xz}^{ 2}+(A_{zz}\pm\omega_{n})^{2}}
\end{equation}
where $\omega_{n}= \gamma_{n}B_{z}$.
Although the ten resolved nuclear spins can only be roughly polarized, it still allows us to individually measure the precession frequencies $f_{\pm}$ of each target nuclear spin while the other nuclear spins stay at a mixed state and do not affect the frequency measurement of the target one. Therefore, the parameters $\{A_{zz},A_{zx}\}$ can be precisely characterized by measuring the precession frequencies individually for each nuclear spin.

We implement an adaptive quantum phase estimation algorithm to efficiently measure these precession frequencies\cite{Griffiths96}.
The basic idea of this adaptive scheme is to perform a sequence of Ramsey interferometry experiments with different precession time $t_{n}=2^{N-n}t_{min}  (n=1,...,N)$  so that the frequency probability distribution is updated step by step and the frequency estimation range is gradually narrowed down.
In each Ramsey experiment, the target nuclear spin is prepared into a superposition state $(|\uparrow\rangle+|\downarrow\rangle)/\sqrt{2}$ and the electron spin is prepared in the $|+1\rangle$ or $|-1\rangle$ state. Subsequently, the target nuclear spin state evolves into $(|\uparrow\rangle+e^{i \phi_{n}}|\downarrow\rangle)/\sqrt{2}$ after $t_{n}$, where $\phi_{n}=2\pi f t_{n}$ carries the information of the to-be-measured frequency $f$.
Before the projective measurement of the nuclear spin along the $x$ axis, a rotation $\hat{R}_{Z}^{\vartheta_{n}}$ along the $z$ axis with an appropriate angle $\vartheta_{n}$ is applied on the nuclear spin to effectively change the measurement basis.
Finally, the nuclear spin is measured with a probability $P_{n} = \frac{1+cos(\phi_{n}-\vartheta_{n})}{2}$ in the $|\uparrow\rangle$ state.
The essential idea of this method is to use $P_{n}$ to update the frequency probability distribution by the Bayesian inference and adaptively change the measurement basis based on the previous outcomes, i.e., to deduce the best rotation angle $\vartheta_{n+1}$ with $\{P_{1},...,P_{n-1}\}$ for minimizing the uncertainty of $P_{n+1}$ through the semiclassical implementation of the quantum phase estimation algorithm \cite{Griffiths96}. After N-step Ramsey experiments, the frequency is estimated to be the value which gives the highest probability.

\begin{figure*}[t]
	\includegraphics[width=150mm]{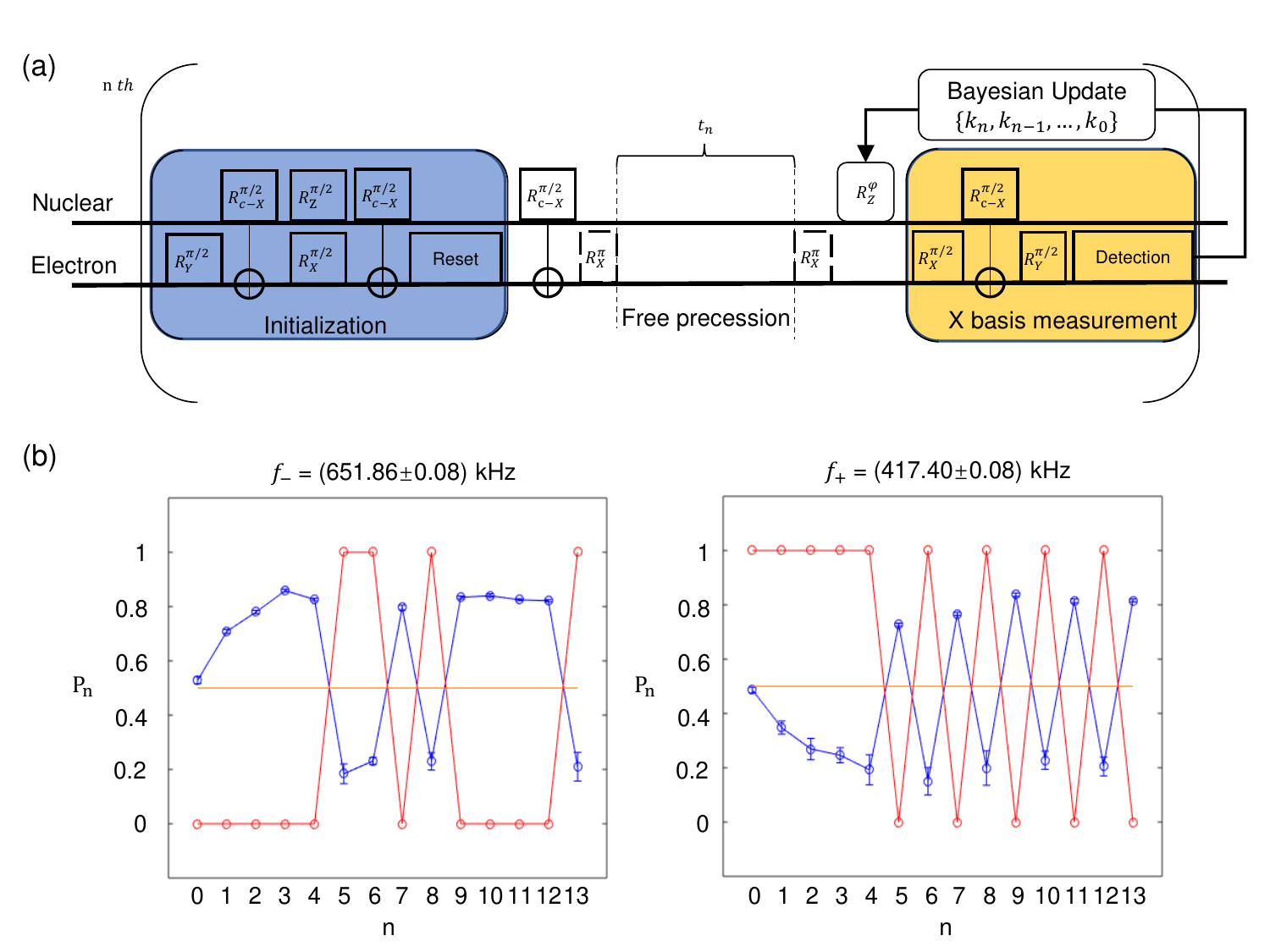}
	\caption{\label{fig:fig2}(a) Experimental sequence to measure the nuclear spin precession frequency under different states of the electron spin through an adaptive method based on the quantum phase estimation algorithm. (b) Frequency measurement results of $f_{-}$ (left panel) and $f_{+}$ (right panel). In the n-th Ramsey experiment, $k_{n}$ (red circle) is estimated by the corresponding outcome of probability $P_{n}$ (blue circles with the error bar). }
\end{figure*}

In the experiment, $t_{min}=$800 ns is determined by the upper bound of the precession frequencies for ensuring $\phi_{n}$ in the window $(0,\pi]$. The maximal time $t_{max} = t_{1}$ is determined by the nuclear spin coherence time which is typically around 10 ms and $t_{max}$ also determines the frequency precision $f_{0} = \frac{1}{2t_{max}}$. The experimental sequence consisting of N-step Ramsey experiments is shown in Fig \ref{fig:fig2}(a). Each Ramsey experiment is repeated $1000$ times. The angle of the Z-rotation gate is updated with $\vartheta_{n+1}= \frac{\vartheta_{n}}{2} + \frac{k_{n}\pi}{2}$ starting with the initial phase $\vartheta_{1}=\frac{\pi}{2}$, where $k_{n}$ equals 0 or 1 depending on $P_{n}>0.5$ or $P_{n}<0.5$, where $k_{n}$ can also be regarded as a binary digit in the form of $f=\sum _{n=0}^{N-1}{2}^{n}\cdot {k}_{i}\cdot {f}_{0} + \varepsilon$ (see Appendix A). According to such a formula, the sequence of Ramsey experiments is implemented to measure the digits in the binary representation of the frequency $f$ one by one, starting from the least significant digit (the last digit).

Fig.\ref{fig:fig2}(c) shows the frequency estimation result as an example. $P_{n}$ for each Ramsey experiment (blue circles) are clearly away from $0.5$ so that $k_{n}$ (red circles) can be estimated correspondingly with a high confidence.
By adaptively changing the measurement basis, $P_{n}$ should approach $0$ or $1$ (move away from $0.5$) in the quantum phase estimation algorithm, but eventually it is limited by the nuclear spin polarization fidelity.
For the first Ramsey experiment, $P_{0}$ is measured to be near $0.5$ because of no adaptive change of the basis before this measurement and a significant decoherence during the long precession time.

Besides the nuclear spin coherence time, another dominant contribution to the systematic error is the magnetic field misalignment.
The transverse magnetic field $B_{x}$ is calibrated to be near zero by minimizing the sum of the two electron spin resonant frequencies $f_{|0\rangle\leftrightarrow |\pm1\rangle}$. However, the intrinsic short coherence time of the electron spin leads to a wide resonance linewidth so that $B_{x}$ can only be calibrated to be smaller than $2.5$ gauss.
The residual $B_{x}$ field can cause a fairly large error in the nuclear spin precession frequency, which is estimate by the form
\begin{equation}
\Delta f_{+} \approx \frac{(A_{zz}\pm\omega_{n})A_{zx}\omega_{ex}}{f_{\pm}(\Delta\pm\omega_{e})}\pm\frac{A_{zx}\omega_{nx}}{f_{\pm}}
\end{equation}
where $\omega_{ex}= \gamma_{e}B_{x}$, $\omega_{nx}= \gamma_{n}B_{x}$.(See details in Appendix D).

\begin{table}
	\caption{Measured nuclear spin hyperfine parameters and initialization fidelities. The number in the bracket denotes the error bar in the last digit. \label{tab:tab1} }
	\centering
	\begin{tabular}{|ll|l|l|l|l|l|}
		\hline		
		& No. 		&1	      &2         &3        &4         &5\\
		\hline
		& $A_{zx}/kHz$ &208(1)   &72(1)    &72(1)      &71(1)   &43(1)\\
		& $A_{zz}/kHz$ &566.0(3) &45.9(1)  &-15.1(1)   &118.1(1) &5.50(7)  \\
		& $F_{ini}/\%$ &95(2)     &94(1)     &93(1)    &97(1)     &92(1)   \\
		\hline
		& No. 		&6         &7        &8        &9        &10\\
		\hline
		& $A_{zx}/kHz$  &33(1)     &32(1)    &31(1)    &29(1)    &17(1)\\
		& $A_{zz}/kHz$  &-49.64(5) &46.34(5) &27.09(5) &28.70(5) &-14.28(3)\\
		& $F_{ini}/\%$  &93(1)     &81(1)    &78(1)    &78(1)    &86(1)\\
		\hline
	\end{tabular}	
	
\end{table}

Hyperfine parameters $A_{zz}, A_{zx}$ of the resolved nuclear spin are calculated by Equation (3) and (4) and shown in TABLE \ref{tab:tab1}.
With the more accurate measurement of the hyperfine parameters, we numerically simulate the dynamical decoupling spectroscopy based on these parameters and compare the simulation results with the experiment data in Fig. \ref{fig:fig1}(c).
The simulation results cover most of the signals in the spectroscopy, but the unresolved nuclear spin bath still leads to some deviation in certain regions.

\section*{A universal set of gates}

With the precisely characterized parameters, single and controlled quantum gates on nuclear spins can be optimized by tunning parameters of dynamical decoupling sequences. We implement a universal set of quantum gates for this 11-spin register, including the single-bit gates for each electron and nuclear spin, and the controlled rotation gates between the electron spin and each nuclear spin. We choose the following three rotations $\hat{R}^{\pi/2}_{X}, \hat{R}^{\pi/2}_{Z}, \hat{R}^{\pi/4}_{Z}$ as a universal set of single-bit gates, where the subscript and the superscript denotes the gate rotation axis and rotation angle, respectively. These gates are performed on each qubit in this spin register. The entangling gate is chosen as $\hat{R}^{\pi/2}_{c-X}$, where the electron spin acts as the control bit, and each nuclear spin undergoes a controlled rotation depending on the electron spin state. To design these gates under constantly on interaction, we numerically optimize the microwave dynamical decoupling sequence taking into account the evolution of all the spins with the precisely calibrated interaction parameters. To model the unresolved nuclear spins in the spin bath, we randomly take 10 additional nuclear spins with interaction parameters uniformly distributed in the range $|A_{zz}, A_{zx}| < 10kHz$ (this range is reasonable as for nuclear spins with larger hyperfine parameters, they should have been identified already) in the simulation in addition to the 10 resolved nuclear spin with well-calibrated interaction parameters.
Nuclear spin single-qubit gates are designed by numerical optimization of the dynamical decoupling sequence that gives a minimal gate duration to mitigate the decoherence effect. The controlled entangling gate
$\hat{R}^{\pi/2}_{c-X}$ is optimized by minimizing the crosstalk to other nuclear spins.

\begin{figure*}[t]
	\includegraphics[width=150mm]{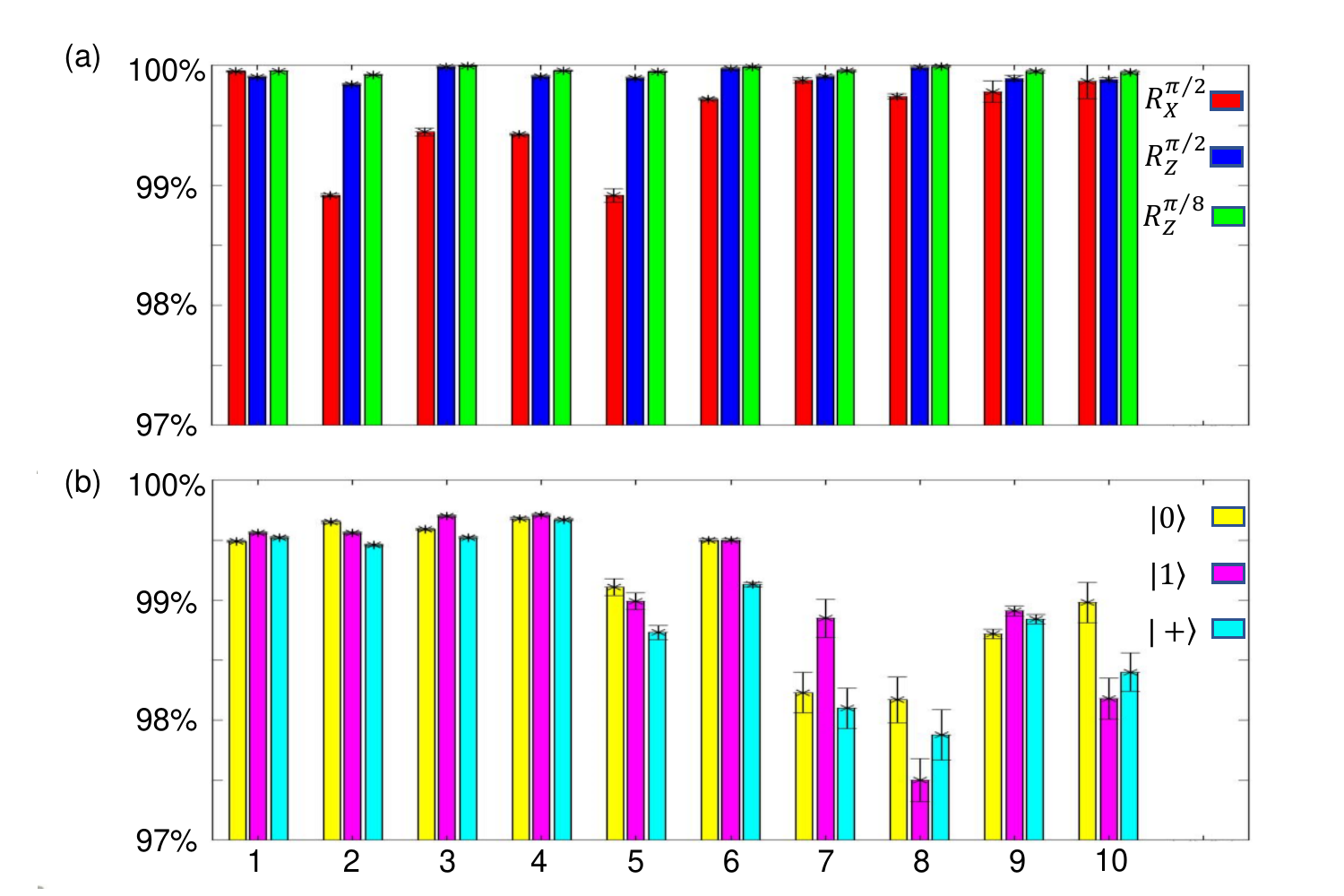}
	\caption{\label{fig:fig3} \textbf{Histogram of estimated gate fidelities for the 10 resolved nuclear spins.} (a) Estimated gate fidelities of single-qubit gates $\hat{R}^{\pi/2}_{X},\hat{R}^{\pi/2}_{Z},\hat{R}^{\pi/4}_{Z}$ for each nuclear spin. (b) The fidelities of the entangling gates $\hat{R}^{\pi/2}_{c-X}$ between the electron spin and each nuclear spin under different initial states $|e \uparrow\rangle$, $|e\rangle=|0\rangle, |1\rangle, |+\rangle$ for the control qubit (electron spin). The gate fidelities are estimated through fit to the decay of the state fidelity under repeated application of the same gates.}
\end{figure*}

In experiment, we use these optimized nuclear spin gates to polarize the resolved nuclear spins with fidelities shown in TABLE.\ref{tab:tab1}. The numbers are significantly improved compared with the case when we have only a rough estimation of the hyperfine interaction parameters through the dynamical decoupling spectroscopy, although for very weakly interacting nuclear spins, the initialization is still limited due to the crosstalk error with the unresolved nuclear spin bath. To characterize the gate errors, we follow a simple calibration method to repeatedly apply the same gates and investigate how the overall state fidelity decays with the number of applied gates \cite{zu2014experimental, rong2015experimental}. The method may over-estimate the gate fidelity as some unitary errors could cancel with each other under repeated application of the gates, however, it gives a rough indicator of the gate fidelity and is doable for this system.
To estimate single-qubit gate error, we polarize nuclear spins to the $|\uparrow\rangle$ state and apply the same gate M times so that the composite operation is an identity and measure the fidelity decay between the outcome state and the ideal target state.
The gate fidelity $F_{gate}$ is deduced by fitting the state fidelity decay $F_{state}(M)$ with a linear function $F_{state} = F_{init} - M\cdot F_{gate}$, where $F_{init}$ is the nuclear spin initialization fidelity.
The electron spin is prepared into the states $|0\rangle$ and $|-1\rangle$, respectively, to eliminate the potential bias in the initialization, and we take the average value of these two cases as the estimated gate fidelity in Fig.\ref{fig:fig3}(a). The gate fidelities estimated by such a method are shown in Fig.\ref{fig:fig3}(a).

Analogous to single-qubit gate fidelity estimation, we also estimate the fidelity of the two-qubit entangling gate $\hat{R}_{c-X}^{\pi/2}$ by measuring the electron-nuclear spin joint state fidelity.
We prepare the electron spin in $|0\rangle, |-1\rangle$ and $|+\rangle=(|0\rangle+ |-1\rangle)/\sqrt2 $ states, respectively, and show the corresponding results in Fig.\ref{fig:fig3}(b). Each nuclear spin is initialized to the $|\uparrow\rangle$ state.
The results indicate that the two-qubit gate $\hat{R}_{c-X}^{\pi/2}$ fidelities of the nuclear spins with strong hyperfine interaction strength are higher than the gate fidelities of the nuclear spins with weaker hyperfine parameters. It implies that the
spin bath still introduce crosstalk error to the gate operations of the resolved nuclear spins and the weakly-coupled nuclear spins suffer more crosstalk to the spin bath than the strongly-coupled ones.

\section*{Summary}

In summary, we have demonstrated experimental Hamiltonian learning in an 11-qubit solid-state quantum spin register with constantly on interaction in a diamond NV center. The learning of the Hamiltonian parameters is implemented by combining the rough parameter estimation from the dynamical decoupling spectroscopy and the precise determination of each parameter through the adaptive measurement of the nuclear spin precession frequency with the semiclassical quantum phase estimation algorithm.  As an example application of the learnt multi-qubit interaction Hamiltonian with precisely determined parameters, we design and optimize a universal set of quantum gates on these 11 spin qubits under the constantly-on interaction and use the knowledge of the learnt interaction parameters to minimize the crosstalk errors. The gate fidelities, estimated from the decay of the state fidelities from the repetitive application of the quantum gates, are quite high, which indicates that the Hamiltonian learning approach is useful to minimize the crosstalk error in the multi-qubit platform. In future, we could implement longer dynamical decoupling sequences to identify other more weakly coupled nuclear spins from the spin bath. This knowledge will help to further improve the initialization and the gate fidelities for those weakly coupled nuclear spins and reduce the crosstalk error between them.   Some of the Hamiltonian learning techniques adopted here, such as the two-step protocol and the adaptive quantum phase estimation algorithm, may also find applications in other multi-qubit systems to characterize the full interaction Hamiltonian and to minimize the crosstalk errors for quantum gates operations.

\textbf{Note Added:} After completion of this work, we became aware of a recent preprint \cite{bradley201910} that demonstrated a universal set of quantum gates in a 10-qubit quantum spin register.

\textbf{Acknowledgements:}
This work was supported by the Ministry of Education of China, Tsinghua University,
and the National key Research and Development Program of China (2016YFA0301902).

\appendix
\section{Appendix A: Frequency measurement by the adaptive quantum phase estimation}
An arbitrary frequency $f$ can be written into a binary form $f=\sum _{i=1}^{N}{2}^{i-1}\cdot {k}_{i}\cdot {f}_{0} + \varepsilon$, where $k_{i}\in\{0,1\}$ and $\varepsilon =\sum _{i=0}^{-\infty}{2}^{i-1}\cdot {k}_{i}\cdot {f}_{0}$ is the uncertainty, $f_{0} = \frac{1}{2^{N}t_{min}}$ as described in the main text.
In the first Ramsey experiment with $t_{1} = 2^{N-1}t_{min}$, the final phase $\phi_{1} = 2\pi ft_{1}$ can be simplified as $\phi_{1} = 2K\pi+k_{1}\pi+\alpha\pi$, where $K$ is an unknown integer determined by $\{k_{2}, k_{3},...,k_{N}\}$ and $\alpha \in (0,1)$ determined by $\varepsilon$. With $\vartheta_{1} = \pi/2$, the probability $$P_{1} = \frac{1+cos(\phi_{1}-\vartheta_{1})}{2} = \frac{1+cos(k_{1}+\alpha-1/2)\pi)}{2}$$ If $k_{1}$=0, $P_{1}=\frac{1+cos(\alpha-1/2)\pi}{2}>0.5$; while if $k_{1}$=1, $P_{1}=\frac{1+cos(\alpha+1/2)\pi}{2}<0.5$. Hence, it gives us a way to estimate $k_{1}$.
In the following Ramsey experiments, $k_{n}$ can be estimated by the same way. While the phase produced by the already-measured digits $\{k_{1},...,k_{n-1}\}$ can be well determined and compensated by updating $\vartheta_{n+1}= \frac{\vartheta_{n}}{2} + \frac{k_{n}\pi}{2}$ to cancel out their influence to the probability $P_{n}$ and reduce the fluctuation of the measurement outcomes.

\section{Appendix B: Sample and experiment setup}
\subsection{Sample description}
\begin{figure}[h]
	\includegraphics[width=150mm]{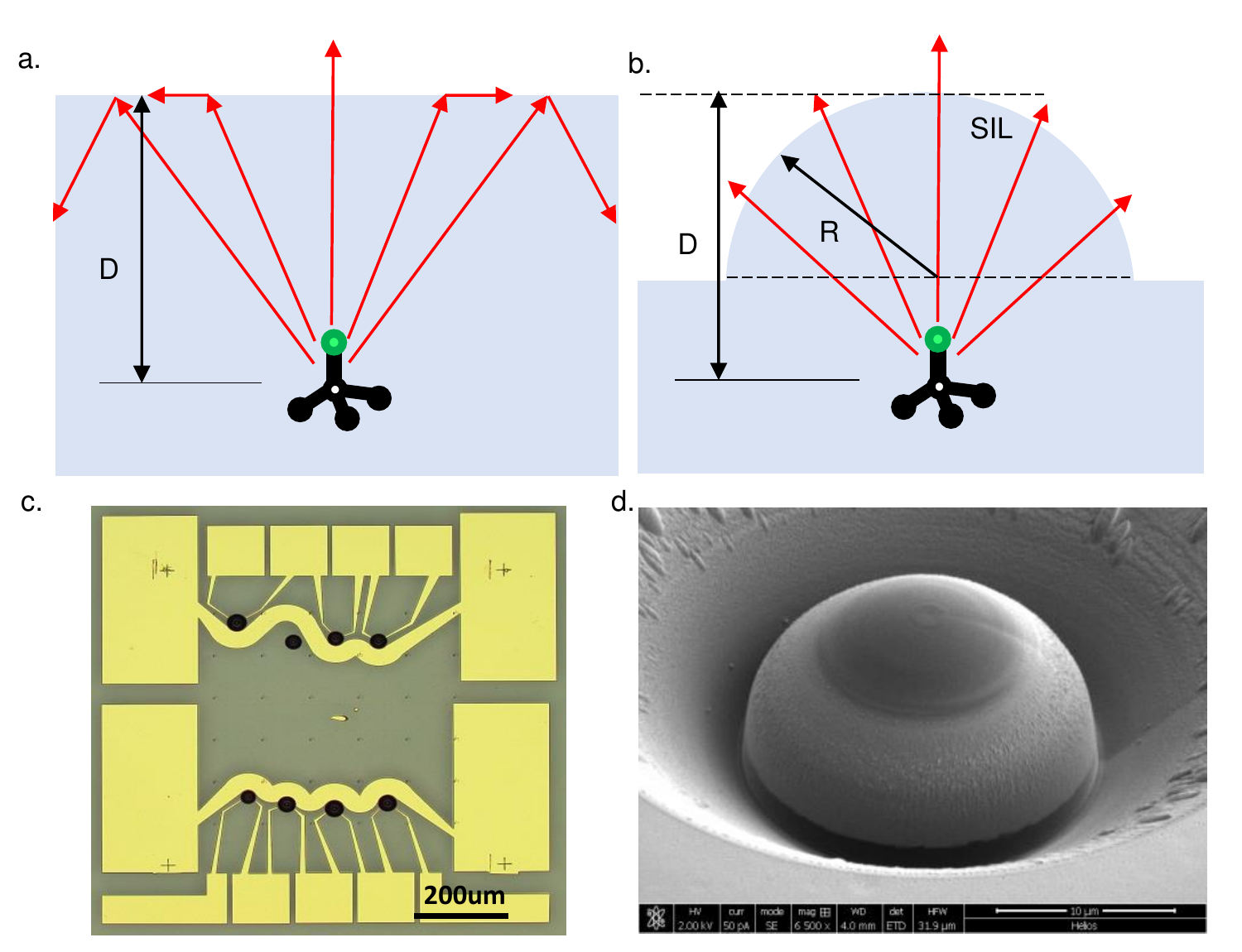}
	\caption{\label{fig:figS1}\textbf{Solid-immersion lens.} (a,b) Sketch of fluorescence emission from a single NV center (a) without or (b) with a solid-immersion len. (c) Image of the sample surface under optical microscope. Each SIL (black circle) is surrounded by a stripline and two electrodes which are connected with two rectangular panels. Four crossings around the corners and black dot arrays are markers pre-fabricated on the surface.
		(d) Image of a SIL under scanning electron microscope.}
\end{figure}

The electronic grade diamond sample used in this work is produced by Element Six with a $\langle 100\rangle$ crystal orientation and a natural abundance ($\sim1.1\%$) of $^{13}C$ atoms.
We fabricate a solid-immersion lens (SIL) on the surface at the site of a single NV center to enhance the fluorescence collection efficiency. The SIL fabrication procedure is following.
We first search individual NV centers at the depth around $7\sim 10$ $\mu m$ under the surface, then obtain their precise coordinates $(x, y, D)$ of each NV center according to the relative position to the references which are the pre-fabricated markers on the surface of the diamond sample, where D is the depth of a single NV center. 
Through numerical simulation, we calculate the optimal SIL radius as D/1.4 to achieve a minimized reflection for the NV fluorescence from the surface (Fig.\ref{fig:figS1}(a,b)). Subsequently, we apply a focused ion beam (FIB) with a hemisphere pattern (Fig.\ref{fig:figS1}.(d)) to fabricate SIL targeting on each NV center. 
For the NV center we use in this work, its fluorescence count rate ($\sim$750kHz) under 532nm laser illumination is enhanced about 7 times by the fabricated SIL.

After fabricating SILs for eight NV centers in Fig.\ref{fig:figS1}(c), we fabricate two 200nm-thickness gold striplines surrounding these SILs (black dots) to delivery the AC current. Each stripline are connected with two panels which are used to wire bonded with the microwave coplanar waveguide.
Two gold electrodes which have a pointing difference of 120 degrees are fabricated for each SIL and used to adjust the local static electric field, although these electrodes are not used in this work.

\subsection{Experiment Setup}
We perform the experiment in a cryogenic temperature ($\sim$8K) high-vacuum ambient provided by a commercial cryostat (Nanoscale Workstation, Montana Instruments).
Inside the cryostat, the sample is mounted on a 3-dimensional positioner from Attocube with sub-micrometer precision to address the SIL. We use a confocal microscopy, with an objective lens with NA =0.95 and a scanning galvo mirror (Thorlabs, GVS212), to address and detect single NV centers.
Outside the cryostat, a permanent magnet is placed on the 3-axis motorized translation stage (PT3-Z8, Thorlabs) to provide an external magnetic field of 495 gauss for the NV centers. We align the magnetic field along the NV-symmetry axis by finding the position where $f=f_{-}+f_{+}$ is minimal ($\sim2877.6MHz$ at 8K), where $f_{\pm}$ are the resonant frequencies of transitions $|0\rangle \leftrightarrow|\pm1\rangle$. 
The readout and initialization of the electron spin are realized by two 637-nm continuous-wave lasers which are frequency-locked by a wavemeter (HighFinesse WS-7). A 532nm green laser is used to ionize the NV center to the negative charge state. We use three acoustic optical modulators (AOM) as optical switches to tailor these laser beams respectively.

Microwave field for coherently controlling NV electron spin is generated by using an IQ-mixer (Marki-microwave IQ-1545) to combine a microwave signal with $1.39 GHz$ ($4.16 GHz$) from a microwave source (Keysight N5181B) and two 100-MHz radio-frequency analog signals from an arbitrary-waveform generator (AWG, Tektronix 5014C). 
We modulate phase and amplitude of the RF signal by programming the AWG to tailor the experiment sequence with 1 ns time resolution. 
After IQ mixer, two amplifiers (Minicircuits ZHL-30W-252-S+ and a home-made amplifier) amplify two generated microwave signals with frequencies $1.49 GHz$ and $4.26 GHz$, respectively. 
We combine two microwave signals and subsequently delivery them to a coplanar waveguide (CPW) which is inside the cryostat and wire-bonded with the surface gold panels. The digital markers of AWG are used to switch the lasers and microwave signals and control the detection windows.

\section{Appendix C: Experiment schematic}

\begin{figure}[h]
	\includegraphics[width=150mm]{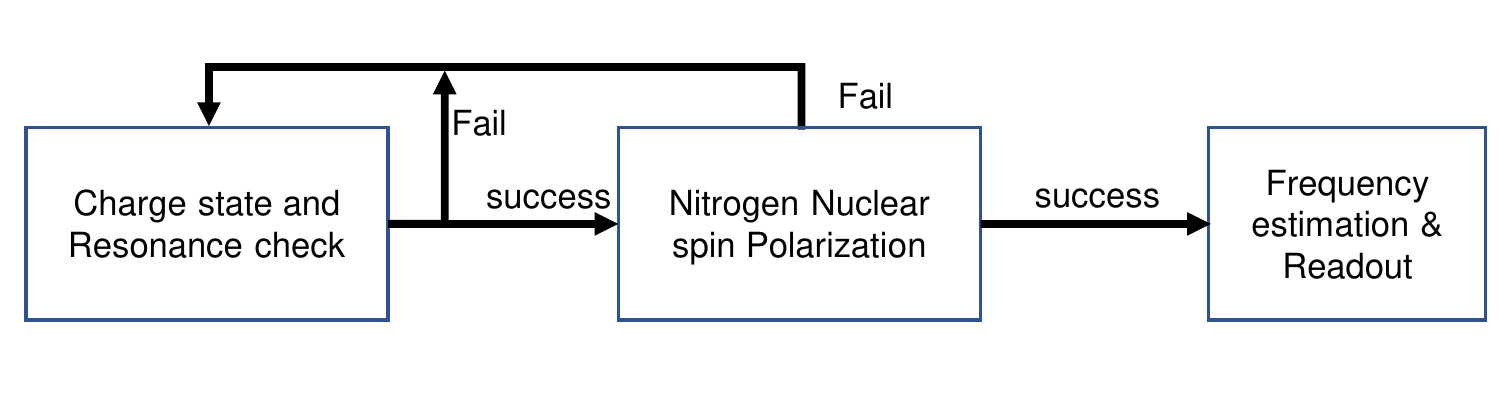}
	\caption{\label{fig:figS2} \textbf{Experiment Schematic}.}
\end{figure}
By employing the technique introduced in Ref.\cite{Lucio11, Cramer16}, we pre-check the charge state and resonance of the NV center with the readout and initialization lasers. 
We open two lasers together for a certain time and record the count of the NV fluorescence. We continue to run the following experiment only in the case that the count is beyond a threshold. 
After the pre-check, we initialize the nitrogen nuclear spin to state $|m_{N}=-1\rangle$ based on measurement\cite{Lucio11}. In the case of success, we subsequently run the experiment sequences.

\section{Appendix D: Hyperfine parameter calculation}

Since the interactions between nuclear spins are negligible in our many-body system, we focus on Hamiltonian of the system composed of one NV electron spin and one nuclear spin. We consider a general form of Hamiltonian by taking into account the full hyperfine interaction tensor $\bf A$ as well as a random external magnetic field $\overrightarrow{B}$. We use Floquet theory to derive zeroth and first order approximation of the non-secular Hamiltonian, thus the full system Hamiltonian is diagonalized in the electron spin subspace. 

Due to the $C_s$ symmetry of the $NV-^{13}C$ system, the hyperfine interaction takes the form: $H_{hfi} = A_{xx}S_{x}I_{x} + A_{yy}S_{y}I_{y} + A_{zz}S_{z}I_{z} + A_{zx} (S_{x}I_{z} + S_{z} I_{x})$, where the Cartesian coordinate $z$ is the NV principle axis, $x$ lies in the plane expanded by the z axis and the target $^{13}C$ atom, $S_{x, y, z} (I_{x, y, z})$ are the spin operators for the electron and nuclear spin respectively. Furthermore, for weakly coupled $^{13}C$ nuclear spins, Fermi contact is ignored, thus we have the constraint $Tr(H_{hfi})=A_{xx}+A_{yy}+A_{zz}=0$. In an arbitrary external magnetic field, the system Hamiltonian can be described by
\begin{equation}
H=\Delta S_{z}^{2}+\omega_{ex}S_{x}+\omega_{ey}S_{y}+\omega_{ez}S_{z}+\omega_{nx}I_{x}+\omega_{ny}I_{y}+\omega_{nz}I_{z}+A_{xx}S_{x}I_{x}+A_{yy}S_{y}I_{y}+A_{zz}S_{z}I_{z}+A_{zx}(S_{z}I_{x}+S_{x}I_{z})
\end{equation}
where $\omega_{e(n)x(y, z)}=\gamma_{e(n)}B_{x(y, z)} $ and $B_{x, y, z}$ is the external magnetic field on the corresponding axis. 

Afterwards, we go to the rotating frame of $U=e^{it(\Delta S_{z}^{2}+\omega_{ez}S_{z})\otimes \mathbb{I}}$, the effective Hamiltonian becomes, 
\begin{equation}
H_{eff}=H_{0}+ e^{i(\Delta+\omega)t}H_{++}+ e^{-i(\Delta+\omega)t}H_{+-}+ e^{i(\Delta-\omega)t}H_{-+}+ e^{-i(\Delta-\omega)t}H_{--}
\end{equation}
where 
\begin{equation}
\begin{split}
H_{0}&=A_{zz}S_{z}I_{z}+A_{zx}S_{z}I_{x}+\omega_{nx}I_{x}+\omega_{ny}I_{y}+\omega_{nz}I_{z}\\
H_{++}&=|+1\rangle\langle0|\otimes \frac{1}{\sqrt{2}}(\omega_{ex}\mathbb{I}-i\omega_{ey}\mathbb{I}+A_{xx}I_{x}-iA_{yy}I_{y}+A_{zx}I_{z})\\
H_{+-}&=|0\rangle\langle+1|\otimes \frac{1}{\sqrt{2}}(\omega_{ex}\mathbb{I}+i\omega_{ey}\mathbb{I}+A_{xx}I_{x}+iA_{yy}I_{y}+A_{zx}I_{z})\\
H_{-+}&=|-1\rangle\langle 0|\otimes \frac{1}{\sqrt{2}}(\omega_{ex}\mathbb{I}+i\omega_{ey}\mathbb{I}+A_{xx}I_{x}+iA_{yy}I_{y}+A_{zx}I_{z})\\
H_{--}&=|0\rangle\langle-1|\otimes \frac{1}{\sqrt{2}}(\omega_{ex}\mathbb{I}-i\omega_{ey}\mathbb{I}+A_{xx}I_{x}-iA_{yy}I_{y}+A_{zx}I_{z})
\end{split}
\end{equation}

Using the Floquet theory\cite{leskes10}, the effective Hamiltonian can be highly approximated by the zeroth and first order approximation. 
\begin{equation}
H_{eff}\approx H_{Floquet}=H^{0}_{eff}+H^{1}_{eff}
\end{equation}
where $H^{0}_{eff}=H_{0}$ and $H^{1}_{eff}=\dfrac{1}{\Delta+\omega_{e}}[H_{++},  H_{+-}]+\dfrac{1}{\Delta-\omega_{e}}[H_{-+}, H_{--}]$.

This Hamiltonian after the Floquet approximation is diagonalized in the electron spin subspace, thus can be rewritten as
\begin{equation}
H_{Floquet}=|+1\rangle\langle+1|\otimes H_{n+}+|0\rangle\langle0|\otimes H_{n0}+|-1\rangle\langle-1|\otimes H_{n-}
\end{equation}
where
\begin{equation}
\begin{split}
H_{n+}&=
\begin{bmatrix}
\frac{A_{zz}}{2}+\frac{\omega_{nz}}{2}+\frac{(A_{xx}-A_{yy})^2+(A_{zx}+2\omega_{ex})^2+4\omega_{ey}^2}{8(\Delta+\omega_{ez})} 
& \frac{A_{zx}}{2}+\frac{\omega_{nx}-i\omega_{ny}}{2}+\frac{2A_{xx}\omega_{ex}+A_{yy}(A_{zx}-2i\omega_{ey})}{4(\Delta+\omega_{ez})} \\
\frac{A_{zx}}{2}+\frac{\omega_{nx}+i\omega_{ny}}{2}+\frac{2A_{xx}\omega_{ex}+A_{yy}(A_{zx}+2i\omega_{ey})}{4(\Delta+\omega_{ez})} & -\frac{A_{zz}}{2}-\frac{\omega_{nz}}{2}+\frac{(A_{xx}+A_{yy})^2+(A_{zx}-2\omega_{ex})^2+4\omega_{ey}^2}{8(\Delta+\omega_{ez})} 
\end{bmatrix}\\
H_{n0}&=
\begin{bmatrix}
\frac{\omega_{nz}}{2}-\frac{\Delta(A_{xx}^{2}+A_{yy}^{2}+(A_{zx}+2\omega_{ex})^2+4\omega_{ey}^2)-2A_{xx}A_{yy}\omega_{ez}}{4(\Delta^{2}-\omega_{e}^{2})} & \frac{\omega_{nx}-i\omega_{ny}}{2}-\frac{2\Delta A_{xx}\omega_{ex}+A_{yy}(A_{zx}\omega_{ez}-2i\Delta\omega_{ey})}{2(\Delta^{2}-\omega_{e}^{2})}  \\
\frac{\omega_{nx}+i\omega_{ny}}{2}-\frac{2\Delta A_{xx}\omega_{ex}+A_{yy}(A_{zx}\omega_{ez}+2i\Delta\omega_{ey})}{2(\Delta^{2}-\omega_{e}^{2})}   & \frac{-\omega_{nz}}{2}-\frac{\Delta(A_{xx}^{2}+A_{yy}^{2}+(A_{zx}-2\omega_{ex})^2+4\omega_{ey}^2)+2A_{xx}A_{yy}\omega_{ez}}{4(\Delta^{2}-\omega_{e}^{2})} 
\end{bmatrix}\\
H_{n-}&=
\begin{bmatrix}
-\frac{A_{zz}}{2}+\frac{\omega_{nz}}{2}+\frac{(A_{xx}+A_{yy})^2+(A_{zx}+2\omega_{ex})^2+4\omega_{ey}^2}{8(\Delta-\omega_{ez})} 
& -\frac{A_{zx}}{2}+\frac{\omega_{nx}-i\omega_{ny}}{2}+\frac{2A_{xx}\omega_{ex}-A_{yy}(A_{zx}+2i\omega_{ey})}{4(\Delta-\omega_{ez})} \\
-\frac{A_{zx}}{2}+\frac{\omega_{nx}+i\omega_{ny}}{2}+\frac{2A_{xx}\omega_{ex}-A_{yy}(A_{zx}-2i\omega_{ey})}{4(\Delta-\omega_{ez})} & \frac{A_{zz}}{2}-\frac{\omega_{nz}}{2}+\frac{(A_{xx}-A_{yy})^2+(A_{zx}-2\omega_{ex})^2+4\omega_{ey}^2}{8(\Delta-\omega_{ez})} 
\end{bmatrix}
\end{split}
\end{equation}

The nuclear precession axis and frequency are calculated by
\begin{equation}
\begin{split}
\overrightarrow{n}_{i}&=(Tr(\sigma_{x}H_{ni}),Tr(\sigma_{y}H_{ni}),Tr(\sigma_{z}H_{ni})), i=+,0,-\\
f_{i}&=||n_i||, i = +,0,-
\end{split}
\end{equation}

Therefore, we are able to get the nuclear precession frequency
\begin{equation}
\begin{split}
f_{+}&=\sqrt{(\omega_{nz}+A_{zz}-\frac{A_{xx}A_{yy}}{2(\Delta+\omega_{ez})}+\frac{A_{zx}\omega_{ex}}{\Delta+\omega_{ez}})^2+(\omega_{nx}+A_{zx}+\frac{A_{zx}A_{yy}}{2(\Delta+\omega_{ez})}+\frac{A_{xx}\omega_{ex}}{\Delta+\omega_{ez}})^2+(\omega_{ny}+\frac{A_{yy}\omega_{ey}}{\Delta+\omega_{ez}})^2}\\
f_{0}&=\sqrt{(\omega_{nz}+\frac{A_{xx}A_{yy}\omega_{ez}}{\Delta^2-\omega_{ez}^2}-\frac{2\Delta A_{zx}\omega_{ex}}{\Delta^2-\omega_{ez}^2})^2+(\omega_{nx}-\frac{A_{zx}A_{yy}\omega_{ez}}{\Delta^2-\omega_{ez}^2}-\frac{2\Delta A_{xx}\omega_{ex}}{\Delta^2-\omega_{ez}^2})^2+(\omega_{ny}-\frac{2\Delta A_{yy}\omega_{ey}}{\Delta^2-\omega_{ez}^2})^2}\\
f_{-}&=\sqrt{(-\omega_{nz}+A_{zz}-\frac{A_{xx}A_{yy}}{2(\Delta-\omega_{ez})}-\frac{A_{zx}\omega_{ex}}{\Delta-\omega_{ez}})^2+(-\omega_{nx}+A_{zx}+\frac{A_{zx}A_{yy}}{2(\Delta-\omega_{ez})}-\frac{A_{xx}\omega_{ex}}{\Delta-\omega_{ez}})^2+(\omega_{ny}+\frac{A_{yy}\omega_{ey}}{\Delta-\omega_{ez}})^2}
\end{split}
\end{equation}

In our experiment, the magnetic field around 495 gauss is near perfectly aligned along $z$ axis so that $B_{z}\gg B_{x},B_{y}$. 
In practical, $B_{x},B_{y}$ can only be bounded by 2.5 gauss, so that $\omega_{ex}, \omega_{ey}$ are still in the level of several $MHz$ which is much larger than the hyperfine strength of weakly coupled nuclear spins.
We also have $\Delta, \omega_{ez} \gg \omega_{ex}, \omega_{ey}$ due to the magnetic field orientation.

For example of $f_{+}$, we can neglect the third term in the first parenthesis, the second and third term in the second parenthesis and the entire third parenthesis, then perform zeroth and first order approximation to derive
\begin{equation}
\begin{split}
f_{+}&\approx\sqrt{(\omega_{nz}+A_{zz}+\frac{A_{zx}\omega_{ex}}{\Delta+\omega_{ez}})^2+(\omega_{nx}+A_{zx})^2} \\
&\approx f_{+,0} + \frac{(A_{zz}+\omega_{n})A_{zx}\omega_{ex}}{f_{+,0}(\Delta+\omega_{e})}+\frac{A_{zx}\omega_{nx}}{f_{+,0}}
\end{split}
\end{equation}
where $f_{+,0} = \sqrt{(\omega_{nz}+A_{zz})^2 + A_{zx}^2}$. So we derive the deviation of $f_{+}$ as
\begin{equation}
\Delta f_{+} = f_{+,0} - f_{+}\approx \frac{(A_{zz}+\omega_{n})A_{zx}\omega_{ex}}{f_{+}(\Delta+\omega_{e})}+\frac{A_{zx}\omega_{nx}}{f_{+}} 
\end{equation}
Analogously, we derive
\begin{equation}
\Delta f_{-} = \frac{(A_{zz}-\omega_{n})A_{zx}\omega_{ex}}{f_{-}(\Delta-\omega_{e})}-\frac{A_{zx}\omega_{nx}}{f_{-}} 
\end{equation}

\newpage
\section{Appendix E: Experiment results}

\begin{table}[h]
	\begin{tabular}{|ll|l|l|l|}
		\hline		
		& No.  &$f_{-}/2\pi$ ($kHz$)   &$f_{+}/2\pi$ ($kHz$)   &$\omega_{n}/2\pi ($kHz$)$   \\
		\hline
		& 1    &1115.49(8)			   &-213.09(8)			   &530.177(4)	\\
		& 2    &580.79(4)			   &489.77(4)			   &530.672(4)	\\
		& 3    &520.25(8)			   &550.35(8)			   &530.657(4)	\\
		& 4    &652.50(4)			   &418.36(4)			   &530.636(4)	\\
		& 5    &537.61(4)			   &526.70(4)			   &530.615(5)	\\
		& 6    &481.83(4)			   &581.02(4)			   &530.597(5)	\\
		& 7    &577.74(4)			   &485.27(4)			   &530.578(5)	\\
		& 8    &558.47(4) 			   &504.42(4)			   &530.655(5)	\\
		& 9    &559.88(4)			   &502.59(4)			   &530.569(5)	\\
		& 10   &516.51(4)			   &545.08(4)			   &530.573(4)	\\
		\hline
	\end{tabular}	
	\caption{\textbf{Frequency measurement result.} $f_{\pm}$ is measured by adaptive Ramsey-interferometry experiment. $\omega_{n}$ is estimated by using ODMR technique to measure the transition frequency $|0\rangle\leftrightarrow |\pm1\rangle$.}
\end{table}

\begin{table}[h]
	\begin{tabular}{|ll|l|l|l|l|l|l|l|l|l|l|}
		\hline		
		& No. 			         &1	     &2     &3     &4     &5     &6    &7     &8     &9    &10\\
		\hline
		& $N_{CX_{\pi/2}} $      &10     &38    &14    &30    &26    &56   &56    &34    &52   &65\\
		& $\tau_{CX_{\pi/2}}$    &290   &7648  &20469 &5487  &30449 &7401 &28438 &29824 &11476&18638\\
		& $N_{X_{\pi/2}} $       &23     &25    &62    &24    &90    &122  &58    &95    &81   &261\\
		& $\tau_{CX_{\pi/2}}$    &3632   &3592  &5704  &1678  &3664  &7484 &3598  &3658  &1964 &3811\\		
		& $N_{Z_{\pi/2}} $       &4      &4     &4     &4     &4     &4    &4     &4     &4    &4\\
		& $\tau_{Z_{\pi/2}}$     &23     &41    &44    &38    &44    &47   &41    &42    &42   &45\\		    	
		& $N_{Z_{\pi/8}} $       &1      &1     &1     &1     &1     &1    &1     &1     &1    &1\\
		& $\tau_{Z_{\pi/8}}$     &23     &41    &44    &38    &44    &47   &41    &42    &42   &45\\	
		\hline
	\end{tabular}	
	\caption{\textbf{Nuclear spin gate parameters}: $\pi$-pulse number N and interpulse delay time $\tau$ (ns).\label{tab:tab2} }
\end{table}

\newpage
\section{Appendix F: Gate fidelity estimation}

\begin{figure}[h]\centering
	\includegraphics[width=150mm]{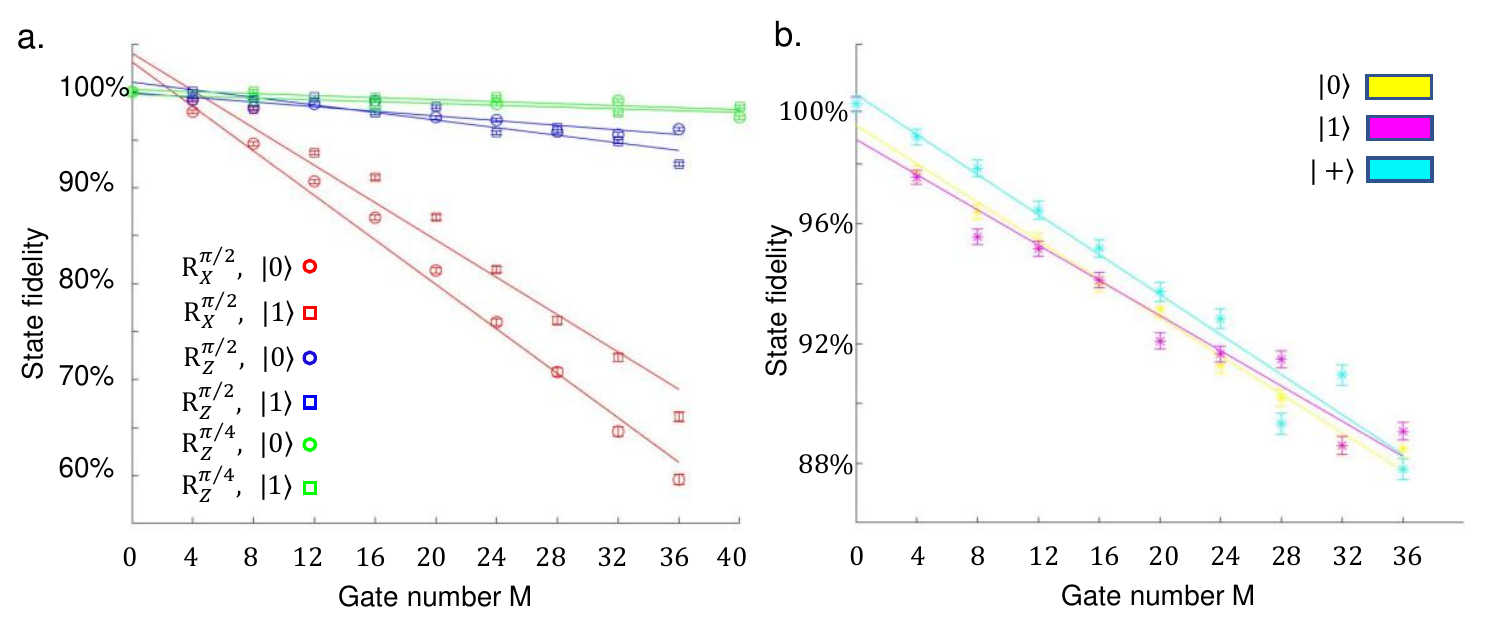}
	\caption{ \label{fig:figS3} \textbf{Gate fidelity estimation.} (a) State fidelity of the target nuclear spin decay as a function of the number of applied single-qubit gates with the initial electron spin state $|0\rangle$ and $|1\rangle$. (b) Fidelity of the joint electron-nuclear spin state decay curve with the initial electron spin state $|0\rangle$, $|1\rangle$ and $|+\rangle$. By fitting state fidelity decay curve with linear function, the slope is taken as gate infidelity.}
\end{figure}

To estimate gate fidelity, we prepare target nuclear spin in $|\uparrow\rangle$ and electron spin in $|0\rangle$ and $|1\rangle$ for single-qubit gates (additional $|+\rangle$ for two-qubit gate). Then we apply the same gate multiple times to make the net operation as an identity depending on the rotation angle. The maximal gate number varies with different gate and nuclear spin because AWG memory limits the total experiment sequence length $l < 32 ms$. The lengths of gate sequences range from tens of nanoseconds to several milliseconds.
Finally, we measure target nuclear spin state fidelity (joint electron-nuclear spin state fidelity) to ideal final state for single-qubit gates (two-qubit gate). To measure the joint electron-nuclear spin state fidelity to ideal state $|0\uparrow\rangle$, we perform ZI, IZ and ZZ measurements to derive the fidelity by the formula $F=(\langle ZI\rangle+\langle IZ\rangle+\langle ZZ\rangle+1)/4$.

\bibliography{ref}

\end{document}